\documentclass[a4paper,aps,prb,twocolumn,showpacs,superscriptaddress,nofootinbib,longbibliography]{revtex4-1}
\usepackage[T1]{fontenc}
\usepackage[utf8]{inputenc}
\usepackage[brazil]{babel}
\usepackage{graphicx}  
\usepackage{dcolumn}   
\usepackage{bm}        
\usepackage{natbib}
\usepackage{latexsym}
\usepackage{mathrsfs}
\usepackage{amssymb}
\usepackage{amsmath}
\usepackage{amscd}
\usepackage{color}
\usepackage{verbatim}
\usepackage{float}

\begin{document}

\title{O experimento de Eratóstenes revisitado}

\author{Levi O. de A. Azevedo}
\affiliation{Instituto de F\'isica, Universidade Federal do Rio de
Janeiro Cx.P. 68.528, 21941-972, Rio de Janeiro-RJ, Brasil}
\author{Orlando S. Ribeiro} 
\affiliation{Secretaria de Educação do Estado do Piauí, 64018-900, Teresina-PI, Brasil}
\affiliation{Secretaria de Educação do Estado do Ceará, 60822-325, Fortaleza-CE, Brasil}
\author{Natanael C. Costa}
\affiliation{Instituto de F\'isica, Universidade Federal do Rio de
Janeiro Cx.P. 68.528, 21941-972, Rio de Janeiro-RJ, Brasil}
\author{Elis H. C. P. Sinnecker} 
\affiliation{Instituto de F\'isica, Universidade Federal do Rio de
Janeiro Cx.P. 68.528, 21941-972, Rio de Janeiro-RJ, Brasil}
\author{Miriam Gandelman} 
\affiliation{Instituto de F\'isica, Universidade Federal do Rio de
Janeiro Cx.P. 68.528, 21941-972, Rio de Janeiro-RJ, Brasil}


\begin{abstract}
Neste trabalho, medimos o raio da Terra reproduzindo o histórico experimento de Eratóstenes, realizado por volta de 240 a.C., nas antigas cidades de Siena e Alexandria.
Aqui, obtemos as medidas nas cidades do Rio de Janeiro-RJ e Teresina-PI, cidades cujas coordenadas de longitude são muito próximas.
Utilizando equipamentos simples, como fios de prumo e cartolinas, medimos, de forma simultânea, a inclinação dos raios solares sobre a superfície da Terra quando o Sol está no ponto mais alto do céu.
Através de dados de satélites -- para a determinação da distância (latitude) entre as cidades -- e pelas medidas obtidas para a diferença dos ângulos em cada cidade, estimamos o raio médio volumétrico da Terra com um erro de $0.5\%$ em relação aos valores da literatura.
Ademais, a partir da diferença do horário em que o Sol fica no ponto mais alto do céu nos locais de medida, estimamos a velocidade angular de rotação da Terra em torno do seu próprio eixo.
Com isso, apresentamos (de forma didática e simples) como obter a curvatura média do planeta, assumindo uma superfície de forma aproximadamente esférica, além da sua velocidade de rotação.
\\
{\bf Palavras-chave: } Experimento de Eratóstenos, raio da Terra, velocidade angular da Terra.
\end{abstract}


\maketitle

\section{Introdu\c{c}\~ao}
\label{Sec:Introduction}

Com a crescente onda de pseudociência e negacionismo, que levantam falsas dúvidas sobre conhecimentos já estabelecidos e testados, é cada vez mais importante a utilização de novas metodologias de ensino e de divulgação científica. 
Neste contexto, a revisitação de experimentos históricos tem um papel crucial, por apresentar de forma didática as etapas da construção do conhecimento.
Isto estimula a divulgação da importância do método científico e da reprodutibilidade de experimentos -- os alicerces da ciência moderna\,\cite{Popper04} -- cujas etapas são indispensáveis para o desenvolvimento de modelos e teorias, assim como suas aplicações em novas tecnologias.

\begin{figure}[t]
\centering{\includegraphics[width=7cm]{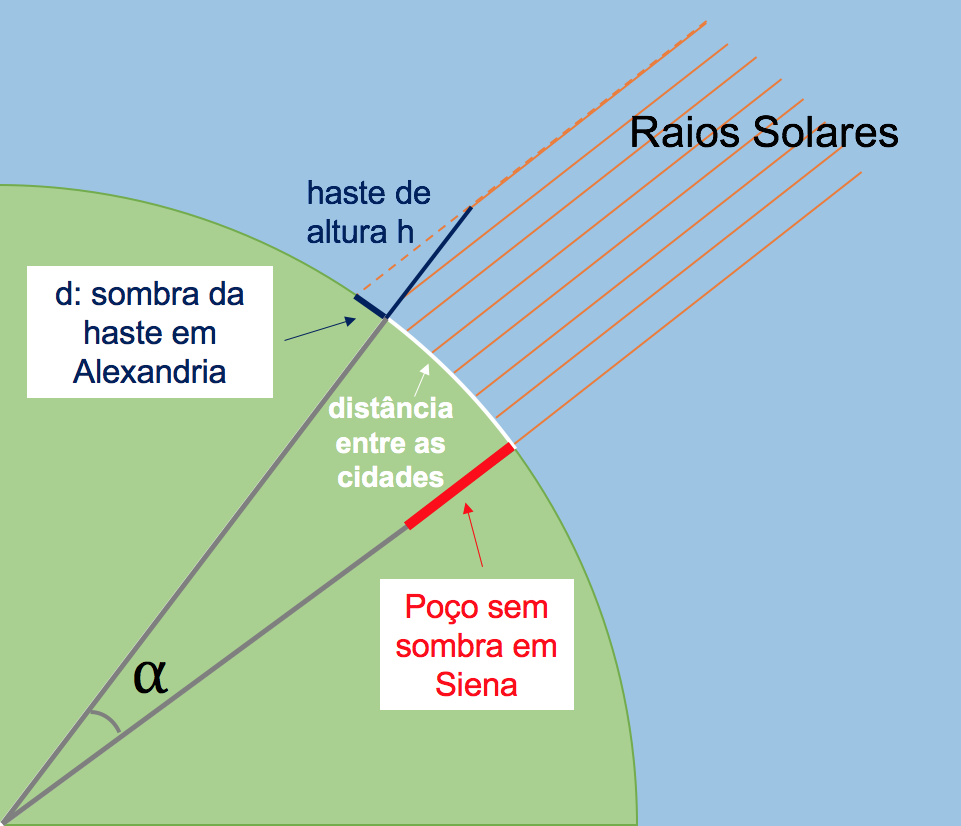}}
\caption{Ilustração representando as duas cidades citadas no texto, Siena e Alexandria, distantes entre si de 5000 estádios (ou 920 km). Figura meramente ilustrativa, com as dimensões da haste e da sombra fora de escala para enfatizar o ângulo, cuja tangente é $\tan(\alpha)=d/h$.}
\label{Fig:fig1}
\end{figure}

Um dos experimentos históricos mais conhecidos é o da medição do raio da Terra feito por Eratóstenes\,\cite{EE01,EE02,EE03,EE04,EE05,EE06,EE07,EE08}, por volta de 240 a.C.
A partir da leitura de documentos presentes na biblioteca de Alexandria, Eratóstenes notou que, no solstício de verão, as paredes dos poços na cidade de Siena não projetavam sombra no fundo dos poços.
Porém, na mesma data e horário, esse fenômeno não acontecia em Alexandria.
Assumindo que os raios solares incidiam paralelamente na Terra, ele levantou a hipótese de que a superfície da Terra deveria possuir uma curvatura, como mostrado na Fig.\,\ref{Fig:fig1}.
Motivado pelas idéias de formas perfeitas e simétricas, elementos comuns nos estudos científicos na Grécia antiga\,\cite{Oliveira16,Freitas21,Pires11}, Eratóstenes propôs que o planeta teria uma forma esférica, e então elaborou um experimento para medir o seu raio.

De forma resumida, o experimento de Eratóstenes consistia em obter uma relação de proporção entre a circunferência da Terra e a distância entre as duas cidades.
Para esta finalidade, ele primeiramente mediu a distância entre Siena e Alexandria, encontrando um valor de cerca de 5.000 estádios (ou cerca de 920 km)\,\cite{newton80,engels85,pinotsis06}.
Em seguida, em dois solstícios de verão (anos) consecutivos, ele mediu a sombra de hastes verticais ao meio-dia, em cada cidade.
Através de relações trigonométricas para a sombra e a altura das hastes, ele estimou o ângulo de incidência dos raios solares em relação à superfície da Terra na cidade de Siena, encontrando cerca de 7,2\textdegree. 
Como exemplificado na Fig.~\ref{Fig:fig1}, e também por argumentos de trigonometria para igualdade de ângulos alternos internos à retas paralelas, Eratóstenes compreendeu que o ângulo de incidência em Siena correspondia, de fato, ao ângulo $\alpha$ entre as duas cidades, a partir do centro da Terra.
Com isso, utilizando a relação de arco de circunferência,
\begin{align}\label{Eq:proporcao}
\frac{\alpha}{360^{\circ}} = \frac{{\rm  dist\hat{a}ncia~entre~as~cidades}}{\rm circunfer\hat{e}ncia~da~Terra}~,
\end{align}
ele estimou a circunferência da Terra em 46.620 km, com cerca de 16\% de erro em relação as estimativas atuais\,\cite{pinotsis06}.
É impressionante notar que, mesmo com um experimento feito há mais de dois milênios, o valor obtido é compatível com as medidas modernas, de 40.030 km\,\cite{Nasa}..

Porém, é importante lembrar que a Terra não é uma esfera perfeita:
a rotação em torno do seu próprio eixo, juntamente com os efeitos da Lua sobre as marés, levam a distorções em relação à uma esfera.
De fato, medidas modernas apontam para uma geometria similar à um elipsóide, com uma diferença entre os raios medidos nos polos, $R_{T (p)}$, em relação ao raio no equador, $R_{T (e)}$ (veja, e.g., as discussões na Ref.\,\onlinecite{Wikipedia} e referências citadas lá).
Apesar disso, estima-se que $|R_{T (e)} - R_{T (p)}| \approx 20$ km, sendo, deste modo, uma diferença desprezível numa análise menos precisa, como a feita por Eratóstenes (e também neste trabalho).
Assim, assumindo uma geometria esférica, medimos o chamado ``raio médio volumétrico'', que é um valor intermediário entre $R_{T (e)}$ e $R_{T (p)}$, e serve como uma primeira estimativa do raio terreste. 
Então, chamamos atenção ao fato que, apesar de desconhecer esses detalhes sobre a Terra, Eratóstenes partiu de uma hipótese correta ao assumir uma forma esférica.
Em outras palavras, o modelo que servia de fundamento para suas pesquisas geográficas (i.e., um planeta esférico) era, com boa precisão, suficiente para explicar os fenômenos e questionamentos da época\,\cite{Hestenes92}.

\begin{figure}[t]
\centering{\includegraphics[width=7cm]{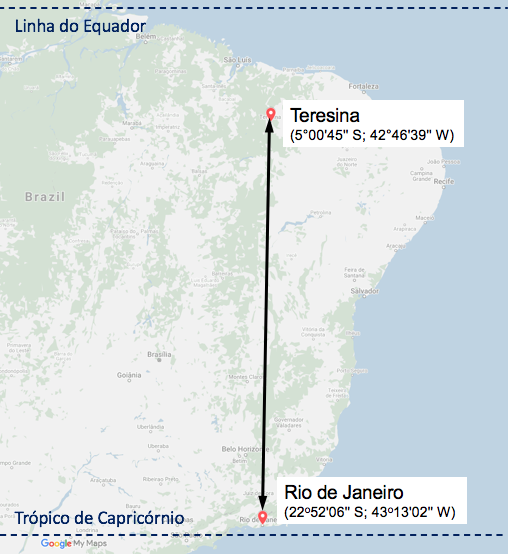} \\ \vspace{0.2cm}
\includegraphics[width=7cm]{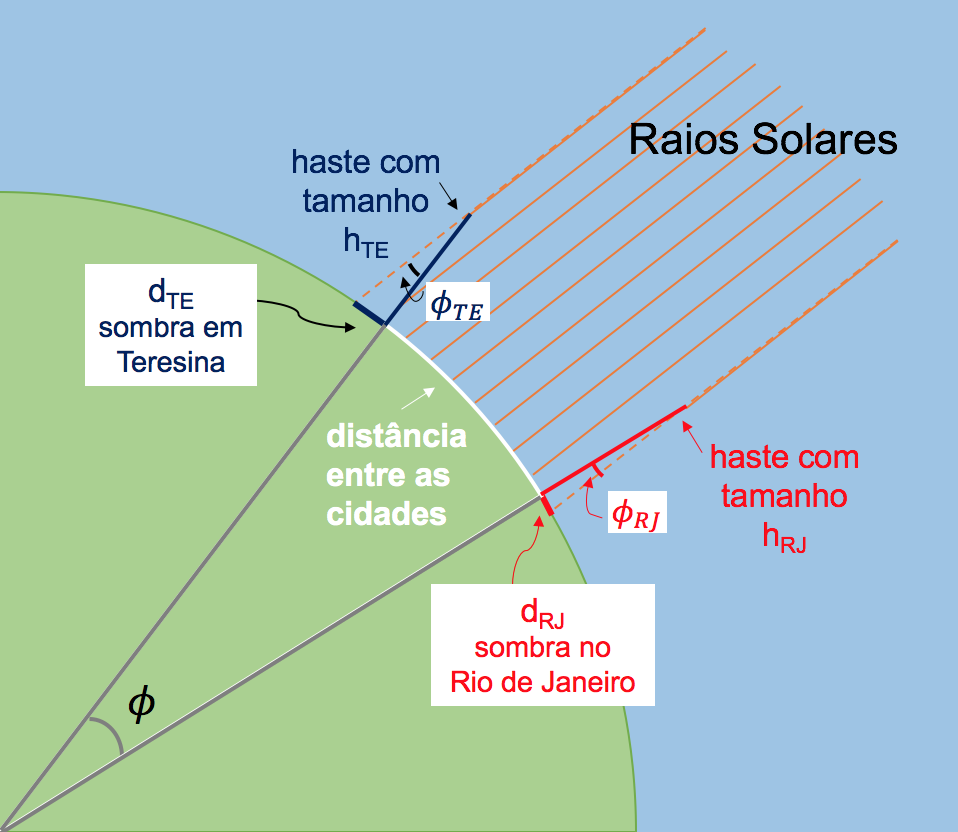}}
\caption{\underline{Painel superior}: Mapa do Brasil com a localização das cidades de Teresina e do Rio de Janeiro.
\underline{Painel inferior}: Ilustração representando o experimento simultâneo feito nas duas cidades. Figura meramente ilustrativa, com as dimensões das sombra fora de escala para enfatizar os ângulos.}
\label{Fig:fig2}
\end{figure}

Em vista desses resultados estimulantes, buscamos neste trabalho reproduzir o experimento de Eratóstenes de forma didática, no âmbito do Museu Interativo da Física da Universidade Federal do Rio de Janeiro (LADIF-UFRJ)\cite{ladif}. O LADIF-UFRJ é um museu científico que se dedica à divulgação de Física e História da Física, com público alvo principal sendo professores e estudantes do ensino fundamental e médio. Nossas medidas foram obtidas nas cidades do Rio de Janeiro-RJ e Teresina-PI, que possuem coordenadas de longitude muito próximas, mas estão muito separadas em coordenadas de latitude (veja, por exemplo, o mapa no painel superior da Fig.~\ref{Fig:fig2}). Por meio de uma análise simples, e usando dados de satélite para a distância entre as cidades, apresentamos como obter o raio médio do planeta, além da sua velocidade de rotação em torno do eixo. Todos os detalhes da metodologia utilizada neste trabalho estão apresentados na próxima seção, enquanto os resultados e as discussões estão na Seção \ref{Sec:Resultados}. Fazemos nossas considerações finais na Seção \ref{Sec:Conclusoes}.

\section{Metodologia}
\label{Sec:Metodologia}

Seguindo o procedimento desenvolvido por Eratóstenes, utilizamos medidas de sombras projetadas no chão, por um objeto posto numa determinada altura.
Para esta finalidade, utilizamos materiais simples, como tubos de plástico extraídos de canetas, fios de prumo, cadeiras, trenas, cartolinas e livros.
É importante destacar que a utilização do fio de prumo, ao invés de uma haste vertical, se dá para reduzir erros associados à inclinação da haste.
Isto é, a atração gravitacional entre a Terra e a massa na ponta do fio de pruno nos garante que o fio estará na vertical durante toda a medida.
Ademais, ressaltamos que o experimento pode ser realizado em qualquer período do ano, independentemente da ocorrência de solstícios.
Deste modo, o valor de $\alpha$ na Eq.\,\eqref{Eq:proporcao} passa a representar a diferença entre os ângulos medidos em cada cidade, como ilustrado no painel inferior da Fig.~\ref{Fig:fig2}.

\begin{figure}[t]
\centering
\includegraphics[scale=0.152]{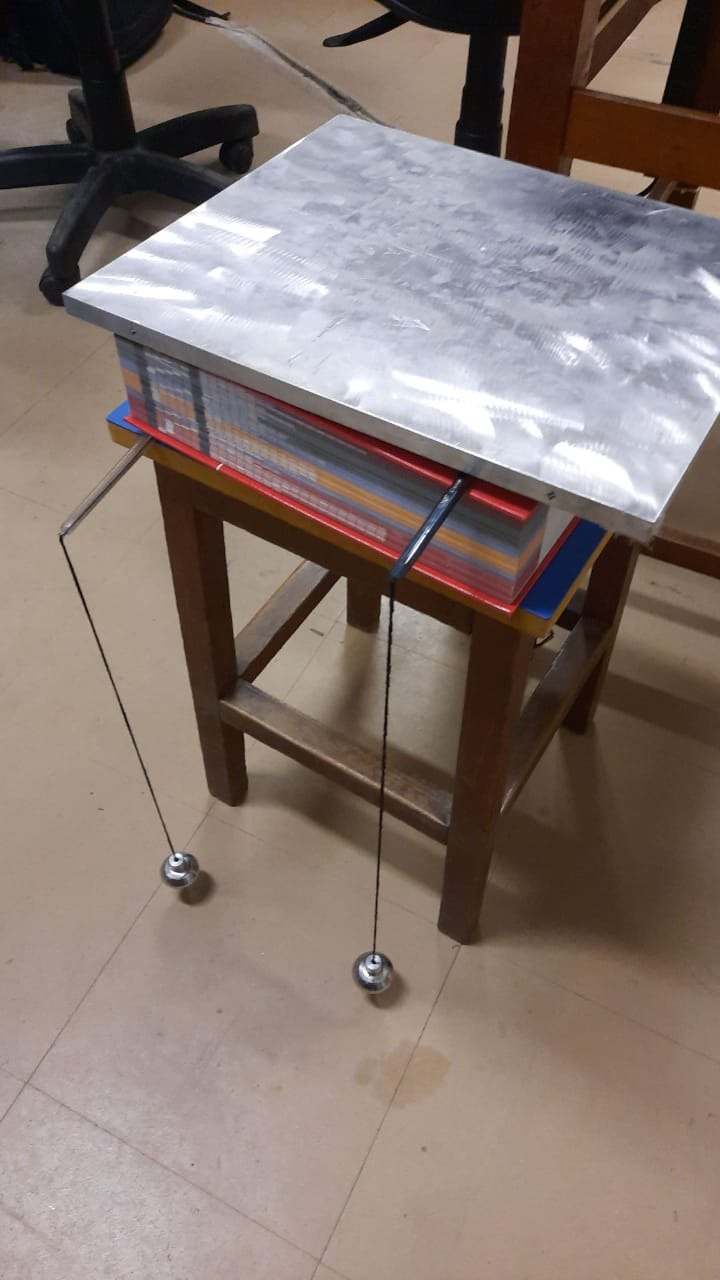}
\includegraphics[scale=0.19]{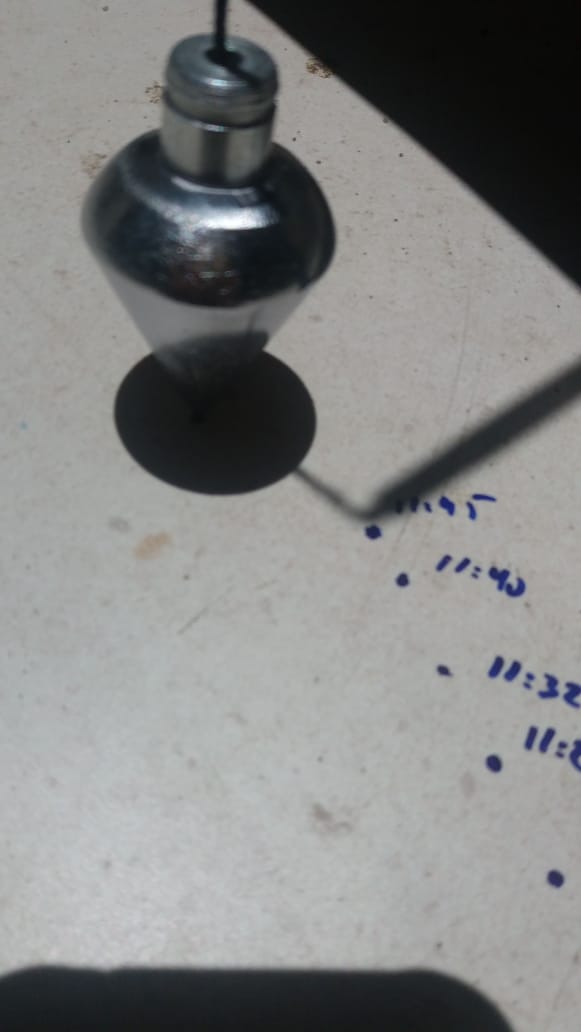}
\caption{\underline{Esquerda}: Um esquema de montagem do aparato utilizado neste trabalho. \underline{Direita}: foto das medidas de projeção das sombras da ponta da caneta feitas no chão, para vários intervalos de tempo.}
\label{Fig:fig3}
\end{figure}

A Figura~\ref{Fig:fig3} apresenta um exemplo de montagem utilizada neste experimento.
Passamos o fio de prumo por dentro do tubo de caneta, que, por sua vez, fica preso aos livros, numa altura que determinamos com a trena.
Mantendo a altura e a posição do tubo de caneta sempre fixas, medimos a distância entre a projeção da ponta do tubo até o eixo do fio de prumo no chão.
Note que a utilização do tubo de caneta aumenta a acurácia da medida do tamanho da sombra do fio, devido à sua espessura fina.
Por fim, como discutido anteriormente, é preciso também determinar a distância entre as cidades.
Para isso, utilizamos ferramentas modernas, de domínio público, como o \textit{Google Earth}, para obter as diferenças/distâncias latitudinais e longitudinais entre Rio de Janeiro e Teresina.

Aqui, chamamos atenção para uma dificuldade inerente ao experimento: devido à distância longitudinal não desprezível entre as duas cidades, os instantes de Sol no ponto mais alto do céu não são simultâneos. 
Além disso, determinar o momento exato em que isso ocorre (i.e., o menor ângulo da projeção da sombra) é um problema não trivial.
Para superarmos essas dificuldades, realizamos várias medidas de tamanho da sombra do obejto, entre as 11h e 13h, com intervalos de 5 a 10 minutos entre elas, como mostrado na Fig.\,\ref{Fig:fig3}.
Isso nos fornece o comportamento do tamanho das projeções de sombra $d$ como função do tempo $t$, como exibido na Fig.\,\ref{Fig:fig4} para ambas as cidades, no qual o valor mínimo $d_{0}$ para o tamanho da sombra pode ser estimado.

Como mostrado no painel direito da Fig.\,\ref{Fig:fig3}, os pontos que marcam a posição da sombra da ponta da caneta ao longo do tempo descrevem, com boa aproximação, uma linha reta.
Deste modo, os tamanhos das sombras $d(t)$ correspondem às distâncias de um ponto (ponta do fio de prumo) a uma reta (as medidas de sombra da ponta da caneta), que, por sua vez, definem uma hipérbole. 
Então, para estimarmos $d_{0}$, fazemos um ajuste não linear dos valores de $d(t)$ pela expressão
\begin{align}\label{Eq:hiperbole}
d(t)=\sqrt{d^{2}_{0} + a^{2}(t-t_{0})^{2}}~,
\end{align}
onde $t_{0}$ é o exato momento em que o mínimo de sombra ocorre, enquanto $a$ é um fator que varia de acordo com a inclinação dos raios solares no dia da medida.
Tal procedimento pode ser feito com auxílio de softwares de análise de dados, como Qtiplot (www.qtiplot.com) ou Root (https://root.cern/).
Esse procedimento nos fornecer valores de $d_{0}$ e $t_{0}$ mais precisos.

Continuamos nossa análise por determinar os ângulos mínimos $\phi^{\rm min}_{R}$ e $\phi^{\rm min}_{T}$ para as cidades do Rio de Janeiro e Teresina, respectivamente.
De posse da altura $h$ da ponta do tubo de caneta, o módulo dos ângulos de incidência dos raios solares em cada cidade é determinado por
\begin{align}\label{Eq:angulo}
|\phi^{\rm min}| = \arctan\bigg(\frac{d_{0}}{h}\bigg)~.
\end{align}
Alertamos que, para a época do ano que as medidas foram realizadas, a projeção mínima de sombra no Rio de Janeiro apontava para o Sul, $\phi^{\rm min}_{R} < 0 $, enquanto em Teresina apontava para o Norte, $\phi^{\rm min}_{T} > 0 $.
Deste modo, sendo $\alpha = \phi^{\rm min}_{T} - \phi^{\rm min}_{R}$ (que neste caso torna-se $\alpha = |\phi^{\rm min}_{T}| + |\phi^{\rm min}_{R}|$), utilizamos a Eq.\,\eqref{Eq:proporcao} para obtenção do raio da Terra.
Ademais, fazendo uso dos valores de $t_{0}$ em cada cidade, também é possível estimar o valor de velocidade rotação $\Bar{\omega}_{T}$ da Terra em torno do seu próprio eixo, como mostrado na próxima Seção.

\section{Resultados}
\label{Sec:Resultados}

\begin{table}[t]
\begin{ruledtabular}
  \begin{tabular}{cccccc}
     & & Rio de Janeiro & & Teresina & \\
    \hline
Início \footnotemark[1]  (GMT-3) & & 11:12:00 \textit{h} & &  11:15:00 \textit{h}  &  \\    
    \hline
 $h$ (mm) & & $507 \pm 2$ & &  $848 \pm 2$ &  \\
    \hline
 $d_{0}$ (mm) & & $111.6 \pm 0.3$  & &  $78.4 \pm 0.8$ & \\
    \hline
$t_{0}$ (min) \footnotemark[2] & & $53.4 \pm 0.2$  & &  $49.0 \pm 0.2$ &  \\
    \hline
$|\phi^{\rm min}|(\rm graus)$ & & $12.42  \pm 0.06$ \footnotemark[3]   & &  $5.3 \pm 0.1$ \footnotemark[4] &  \\
  \end{tabular}
  \end{ruledtabular}
\footnotetext[1]{A diferença se deu por questões metereológicas.}
\footnotetext[2]{Relativo aos seus respectivos inícios de obtenções de medidas. Vide primeiro item desta tabela.}
\footnotetext[3]{Direcionado para o Sul.}
\footnotetext[4]{Direcionado para o Norte.}
    \caption{Dados obtidos no experimento, referentes (i) ao início das medidas em cada localidade, (ii) à altura $h$ da ponta do tudo de caneta, (iii) ao tamanho de projeção mínimo de sombra $d_{0}$, (iv) ao tempo $t_{0}$ transcorrido até o ponto de projeção mínima e (v) ao ângulo de incidência $|\phi^{\rm min}|$.\label{Table1}}
\end{table}

O experimento foi realizado no dia 20 de fevereiro de 2021, a partir de medidas obtidas simultaneamente nas cidades do Rio de Janeiro-RJ e Teresina-PI.
Como primeira passo, obtemos os dados de satélite via plataforma \textit{Google Earth}.
Tomando as coordenadas dos locais de medida como referências, encontramos diferenças latitudinais e longitudinais iguais a 17°51'21'' ($\approx$ 17.86°) e 0°26'23'' ($\approx$ 0.44°), respectivamente, que são equivalentes à distâncias de 1975.8 km (latitude) e 48.8 km (longitude). Aqui, assumimos o erro dos dados de satélite como sendo 0.1 km.

\begin{figure}[t]
\centering
\includegraphics[scale=0.6]{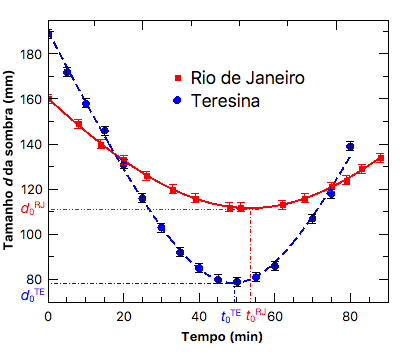}
\caption{Tamanho das projeções de sombra em relação ao fio de prumo (símbolos) medido como fun\c c\~ao do tempo no Rio de Janeiro (curva vermelho sólida) e em Teresina (curva azul tracejada), nas proximidades do meio-dia, em 20 de Fevereiro de 2021. As curvas são ajustes não lineares por uma função hiperbólica.}
\label{Fig:fig4}
\end{figure}

Em seguida, passamos a medir as projeções de sombra.
Essas medidas foram realizadas aproximadamente entre 11:10h e 13:00h, com as marcações sendo feitas em intervalos de 5 a 10 minutos, em ambas as cidades.
Os resultados para $d(t)$ estão apresentados na Fig.\,\ref{Fig:fig4}, enquanto as alturas $h$ das pontas de tubo de caneta estão disponíveis na Tabela \ref{Table1}.
Repetindo o procedimento descrito na seção anterior, obtemos o valor de projeção mínima de sombra através do ajuste hiperbólico dos dados, Eq.\,\eqref{Eq:hiperbole}, como representado nas curvas sólidas da Fig.\,\ref{Fig:fig4}.
De posse de $d_{0}$ e $h$, calculamos o módulo dos ângulos (mínimos) de incidência $|\phi^{\rm min}|$, Eq.\,\eqref{Eq:angulo}, cujos resultados são apresentados na Tabela \ref{Table1}.
Consequentemente, encontramos o ângulo de abertura de latitide $\alpha = (17.7 \pm 0.1)^{\circ}$.
Aqui, como ``controle de qualidade'', é importante notar que esse resultado para $\alpha$ é consistente com os dados de satélite usados como referência, que apontam uma diferença de aproximadamente 17.86° entre os locais de medidas.

Assim como feito por Eratóstenes, faremos aqui uso de argumentos trigonométricos para obter o raio da Terra.
Usando a Eq.\,\eqref{Eq:proporcao}, e fixando a distância latitudinal entre os locais de medida como $1975.8 \pm 0.1$ km, encontramos que a circunferência da Terra é $C_{T}=(40.2 \pm 0.2)\times 10^3 $ km, com as barras de erro sendo obtidas por cálculos usuais de propagação de incerteza de medidas indiretas.
Logo, a nossa estimativa para o raio médio volumétrico terrestre é 
\begin{align}\label{Eq:raioExp}
R_{T}=\frac{C_{T}}{2 \pi} = (6.40 \pm 0.03) \times 10^3 ~~{\rm km.}
\end{align}
Este valor de raio da Terra deve ser comparado com valores da literatura, fornecidos pelo \textit{Space Science Data Coordinated Archive} da agência espacial norte-americana \textit{National Aeronautics and Space Administration} (NASA)\,\cite{Nasa}, de $R_{T {\rm (best)}} = 6371$ km.
Assim, nossos resultados experimentais para $R_{T}$ estão em excelente acordo com a literatura mais recente, com um erro relativo de
\begin{align}
\epsilon_{R} =\frac{|R_{T {\rm (best)}} - R_{T}|}{R_{T {\rm (best)}}} \approx 0.5~\%.
\end{align}

Por fim, vamos examinar a velocidade média de rotação da Terra, $\Bar{\omega}_{T}$.
Para isso, devemos medir a taxa de variação do ângulo de longitude $\Delta\theta$ entre as duas cidades como função do tempo, usando os instantes de projeção mínima de sombra como referências.
De acordo com a Tabela \ref{Table1}, os instantes de Sol no ponto mais alto de céu ocorreram as 12h 04min 00s $\pm 12$ s no local de medida em Teresina, e as 12h 05min 24s $\pm 12$ s no Rio de Janeiro.
Daí, a diferença real de tempo entre a ocorrência de $d_{0}$ em Teresina e Rio de Janeiro é $\Delta t = (84 \pm 17)$ s.
Como já mencionado, por meio dos dados fornecidos por satélite, obtemos a diferença de coordenadas longitudinais de aproximadamente 0°26'23'' $\pm$ 0°00'06'' (barra de erro compatível com as incertezas de $\pm$ 0.1 km de distância de cada local de medida).
Isto equivale a $\Delta\theta = (0.440 \pm 0.002)^{\circ} = (7.68 \pm 0.03) \times 10^{-3}$ rad, e no conduz a
\begin{align}\label{Eq:rotacaoExp}
\nonumber
\Bar{\omega}_{T}=\frac{\Delta\theta}{\Delta t} & = (9 \pm 2) \times 10^{-5} ~~{\rm rad/s}~, \\
& = (0.33 \pm 0.07) ~~{\rm rad/h.}
\end{align}
Novamente, é necessário compararmos esse resultado com os valores da literatura.
De acordo com a Ref.\,\onlinecite{Nasa}, o período médio de rotação da Terra é $T_{T {\rm (best)}} = 23.9345$ h, que leva à $\Bar{\omega}_{T {\rm (best)}}  \approx 0.2625$ rad/h.
Apesar de satisfatório, nosso resultado para $\Bar{\omega}_{T}$ é pouco acurado, por apresentar um erro relativo de $\epsilon_{\omega} \approx 26~\%$.

É importante enfatizar que esse valor de $\epsilon_{\omega}$ não pode ser associado à uma possível flutuação (em relação à média) no período de rotação da Terra, uma vez que a ordem de grandeza dessas variações é de $10^{-3}$ s, e podem ser desprezadas\,\cite{dickey95,gross07}.
De fato, tal erro é devido à imprecisões sistemáticas e de instrumento que se acumulam ao se realizar as medidas de tempo (feitas por relógios usuais) e, consequentemente, levam a incertezas na diferença de tempo da mesma ordem de grandeza de $\Delta t$.
Há pelo menos duas maneiras de superar essa dificuldade sem a necessidade de relógios de alta precisão: (i) realizar medidas em localidades mais separadas longitudinalmente, ou (ii) realizar muitas medidas similares para essas mesmas cidades (em dias diferentes) e, em seguida, fazer uma análise estatística dos resultados.
Em particular, essa última opção eliminaria incertezas aleatórias, e seria tão mais precisa quanto maior fosse o número de medidas realizadas.
Contudo, uma análise estatística vai além do escopo dos objetivos iniciais dessa proposta.

\section{Conclusões}
\label{Sec:Conclusoes}

Neste trabalho, revisitamos o experimento de Eratóstenes para medição do raio médio da Terra, por meio do uso de materiais simples, e de plataformas de acesso público, como o \textit{Google Earth}.
Apresentamos um roteiro para a obtenção e análise dos dados, que nos forneceram (com boa precisão) os ângulos mínimos de incidência e seus respectivos tempos de ocorrência, em cada localidade. De posse desses resultados, resumidos na Tabela \ref{Table1}, encontramos o raio médio volumétrico da Terra como sendo $R_{T}= (6.40 \pm 0.03) \times 10^3 ~~{\rm km}$, em excelente acordo com os dados da literatura, com um erro relativo de apenas 0.5\%.
Similarmente, medimos a velocidade média de rotação da Terra, Eq.\,\eqref{Eq:rotacaoExp}, encontrando um resultado compatível com o da literatura, porém com um erro relativo de 26\%.
Neste caso, explicamos os motivos para isso e propomos alternativas de redução de tal incerteza.
Em linhas gerais, é admirável que um experimento simples, feito com materias caseiros, nos forneça resultados com alto grau de concordância com os valores mais recentes da NASA\,\cite{Nasa}, em particular para o raio da Terra.
Ademais, vale ressaltar que a realização de medidas em dias diferentes permite a observação também da taxa de variação do ângulo de incidência $\phi^{\rm min}$ e, eventualmente, da inclinação do eixo de rotação da Terra (que é de aproximadamente 23.5°).
Mas, deixaremos esses estudos para trabalhos futuros.

Muito além da acurácia dos resultados obtidos, a revisitação deste experimento tem um grande potencial didático, uma vez que serve como uma utilização direta do método científico.
Dentre outras coisas, este experimento tem o potencial de ensinar/abordar técnicas de medida, montagem de aparato experimental, análise dos dados e comparação com resultados da literatura; habilidades que quando desenvolvidas corretamente ajudam a eliminar vícios comuns no aprendizado de ciências naturais.
Assim, esperamos motivar professores e tutores do ensino básico para a utilização deste experimento como ferramenta de desmistificação de ciências e do método científico.

\color{black}
\section{Agradecimentos}
\label{agra}
Os autores agradecem o suporte financeiro das agências brasileiras de financiamento CNPq, CAPES e FAPERJ.
Os autores são gratos ao Prof.\,J.P.~de Lima, Prof.\,F.~Nogueira Lima e ao Dr.\,J.M.~Dias pelas discussões e sugestões durante o desenvolvimento do trabalho.

\bibliography{ref.bib}

\begin{thebibliography}{21}%
\makeatletter
\providecommand \@ifxundefined [1]{%
 \@ifx{#1\undefined}
}%
\providecommand \@ifnum [1]{%
 \ifnum #1\expandafter \@firstoftwo
 \else \expandafter \@secondoftwo
 \fi
}%
\providecommand \@ifx [1]{%
 \ifx #1\expandafter \@firstoftwo
 \else \expandafter \@secondoftwo
 \fi
}%
\providecommand \natexlab [1]{#1}%
\providecommand \enquote  [1]{``#1''}%
\providecommand \bibnamefont  [1]{#1}%
\providecommand \bibfnamefont [1]{#1}%
\providecommand \citenamefont [1]{#1}%
\providecommand \href@noop [0]{\@secondoftwo}%
\providecommand \href [0]{\begingroup \@sanitize@url \@href}%
\providecommand \@href[1]{\@@startlink{#1}\@@href}%
\providecommand \@@href[1]{\endgroup#1\@@endlink}%
\providecommand \@sanitize@url [0]{\catcode `\\12\catcode `\$12\catcode
  `\&12\catcode `\#12\catcode `\^12\catcode `\_12\catcode `\%12\relax}%
\providecommand \@@startlink[1]{}%
\providecommand \@@endlink[0]{}%
\providecommand \url  [0]{\begingroup\@sanitize@url \@url }%
\providecommand \@url [1]{\endgroup\@href {#1}{\urlprefix }}%
\providecommand \urlprefix  [0]{URL }%
\providecommand \Eprint [0]{\href }%
\providecommand \doibase [0]{http://dx.doi.org/}%
\providecommand \selectlanguage [0]{\@gobble}%
\providecommand \bibinfo  [0]{\@secondoftwo}%
\providecommand \bibfield  [0]{\@secondoftwo}%
\providecommand \translation [1]{[#1]}%
\providecommand \BibitemOpen [0]{}%
\providecommand \bibitemStop [0]{}%
\providecommand \bibitemNoStop [0]{.\EOS\space}%
\providecommand \EOS [0]{\spacefactor3000\relax}%
\providecommand \BibitemShut  [1]{\csname bibitem#1\endcsname}%
\let\auto@bib@innerbib\@empty
\bibitem [{\citenamefont {Popper}(2004)}]{Popper04}%
  \BibitemOpen
  \bibfield  {author} {\bibinfo {author} {\bibfnamefont {Karl~R.}\ \bibnamefont
  {Popper}},\ }\href@noop {} {\emph {\bibinfo {title} {A l{\'o}gica da pesquisa
  cient{\'\i}fica}}}\ (\bibinfo  {publisher} {Editora Cultrix, São Paulo,
  Brasil},\ \bibinfo {year} {2004})\BibitemShut {NoStop}%
\bibitem [{\citenamefont {{American Physical Society}}(2006)}]{EE01}%
  \BibitemOpen
  \bibfield  {author} {\bibinfo {author} {\bibnamefont {{American Physical
  Society}}},\ }\href
  {https://www.aps.org/publications/apsnews/200606/history.cfm} {\enquote
  {\bibinfo {title} {{This Month in Physics History: June, ca. 240 B.C.
  Eratosthenes Measures the Earth}},}\ } (\bibinfo {year} {2006})\BibitemShut
  {NoStop}%
\bibitem [{\citenamefont {Weir}(1931)}]{EE02}%
  \BibitemOpen
  \bibfield  {author} {\bibinfo {author} {\bibfnamefont {James}\ \bibnamefont
  {Weir}},\ }\bibfield  {title} {\enquote {\bibinfo {title} {{The method of
  Eratosthenes}},}\ }\href@noop {} {\bibfield  {journal} {\bibinfo  {journal}
  {Journal of the Royal Astronomical Society of Canada}\ }\textbf {\bibinfo
  {volume} {25}},\ \bibinfo {pages} {294} (\bibinfo {year} {1931})}\BibitemShut
  {NoStop}%
\bibitem [{\citenamefont {Longhorn}\ and\ \citenamefont {Hughes}(2015)}]{EE03}%
  \BibitemOpen
  \bibfield  {author} {\bibinfo {author} {\bibfnamefont {Morgana}\ \bibnamefont
  {Longhorn}}\ and\ \bibinfo {author} {\bibfnamefont {Stephen}\ \bibnamefont
  {Hughes}},\ }\bibfield  {title} {\enquote {\bibinfo {title} {{Modern
  replication of Eratosthenes' measurement of the circumference of Earth}},}\
  }\href {\doibase 10.1088/0031-9120/50/2/175} {\bibfield  {journal} {\bibinfo
  {journal} {Physics Education}\ }\textbf {\bibinfo {volume} {50}},\ \bibinfo
  {pages} {175--178} (\bibinfo {year} {2015})}\BibitemShut {NoStop}%
\bibitem [{\citenamefont {Bo{\v{z}}i{\'{c}}}\ \emph {et~al.}(2016)\citenamefont
  {Bo{\v{z}}i{\'{c}}}, \citenamefont {Vu{\v{s}}kovi{\'{c}}}, \citenamefont
  {Popovi{\'{c}}}, \citenamefont {Popovi{\'{c}}},\ and\ \citenamefont
  {Markovi{\'{c}}-Topalovi{\'{c}}}}]{EE04}%
  \BibitemOpen
  \bibfield  {author} {\bibinfo {author} {\bibfnamefont {Mirjana}\ \bibnamefont
  {Bo{\v{z}}i{\'{c}}}}, \bibinfo {author} {\bibfnamefont {Leposava}\
  \bibnamefont {Vu{\v{s}}kovi{\'{c}}}}, \bibinfo {author} {\bibfnamefont
  {Svetozar}\ \bibnamefont {Popovi{\'{c}}}}, \bibinfo {author} {\bibfnamefont
  {Jelena}\ \bibnamefont {Popovi{\'{c}}}}, \ and\ \bibinfo {author}
  {\bibfnamefont {Tatjana}\ \bibnamefont {Markovi{\'{c}}-Topalovi{\'{c}}}},\
  }\bibfield  {title} {\enquote {\bibinfo {title} {{Visualization on the Day
  Night Year Globe}},}\ }\href {\doibase 10.1088/0143-0807/37/6/065801}
  {\bibfield  {journal} {\bibinfo  {journal} {European Journal of Physics}\
  }\textbf {\bibinfo {volume} {37}},\ \bibinfo {pages} {065801} (\bibinfo
  {year} {2016})}\BibitemShut {NoStop}%
\bibitem [{\citenamefont {Babovi{\'{c}}}\ and\ \citenamefont
  {Babovi{\'{c}}}(2014)}]{EE05}%
  \BibitemOpen
  \bibfield  {author} {\bibinfo {author} {\bibfnamefont {Vukota}\ \bibnamefont
  {Babovi{\'{c}}}}\ and\ \bibinfo {author} {\bibfnamefont {Milo{\v{s}}}\
  \bibnamefont {Babovi{\'{c}}}},\ }\bibfield  {title} {\enquote {\bibinfo
  {title} {{The Sun lightens and enlightens: high noon shadow measurements}},}\
  }\href {\doibase 10.1088/0143-0807/35/6/065005} {\bibfield  {journal}
  {\bibinfo  {journal} {European Journal of Physics}\ }\textbf {\bibinfo
  {volume} {35}},\ \bibinfo {pages} {065005} (\bibinfo {year}
  {2014})}\BibitemShut {NoStop}%
\bibitem [{\citenamefont {Pereira}(2006)}]{EE06}%
  \BibitemOpen
  \bibfield  {author} {\bibinfo {author} {\bibfnamefont {Paulo Cesar~R}\
  \bibnamefont {Pereira}},\ }\bibfield  {title} {\enquote {\bibinfo {title}
  {Revivendo {E}rat{\'o}stenes},}\ }\href@noop {} {\bibfield  {journal}
  {\bibinfo  {journal} {Revista Latino-Americana de Educa{\c{c}}{\~a}o em
  Astronomia}\ ,\ \bibinfo {pages} {19--38}} (\bibinfo {year}
  {2006})}\BibitemShut {NoStop}%
\bibitem [{\citenamefont {{Sociedade Astronômica Brasileira}}(2021)}]{EE07}%
  \BibitemOpen
  \bibfield  {author} {\bibinfo {author} {\bibnamefont {{Sociedade Astronômica
  Brasileira}}},\ }\href
  {https://sites.google.com/site/projetoerato/get-started} {\enquote {\bibinfo
  {title} {{Projeto Eratóstenes Brasil}},}\ } (\bibinfo {year}
  {2021})\BibitemShut {NoStop}%
\bibitem [{\citenamefont {Santos}\ \emph {et~al.}(2012)\citenamefont {Santos},
  \citenamefont {Voelzke},\ and\ \citenamefont {Ara{\'u}jo}}]{EE08}%
  \BibitemOpen
  \bibfield  {author} {\bibinfo {author} {\bibfnamefont {Ant{\^o}nio Jos{\'e}
  de~Jesus}\ \bibnamefont {Santos}}, \bibinfo {author} {\bibfnamefont
  {Marcos~Rincon}\ \bibnamefont {Voelzke}}, \ and\ \bibinfo {author}
  {\bibfnamefont {Mauro S{\'e}rgio Teixeira~de}\ \bibnamefont {Ara{\'u}jo}},\
  }\bibfield  {title} {\enquote {\bibinfo {title} {O projeto {E}rat{\'o}stenes
  a reprodu{\c{c}}{\~a}o de um experimento hist{\'o}rico como recurso para a
  inser{\c{c}}{\~a}o de conceitos da {A}stronomia no ensino m{\'e}dio},}\
  }\href@noop {} {\  (\bibinfo {year} {2012})}\BibitemShut {NoStop}%
\bibitem [{\citenamefont {De~Oliveira}\ \emph {et~al.}(2016)\citenamefont
  {De~Oliveira}, \citenamefont {Lima},\ and\ \citenamefont
  {Bertuola}}]{Oliveira16}%
  \BibitemOpen
  \bibfield  {author} {\bibinfo {author} {\bibfnamefont {TB}~\bibnamefont
  {De~Oliveira}}, \bibinfo {author} {\bibfnamefont {VT}~\bibnamefont {Lima}}, \
  and\ \bibinfo {author} {\bibfnamefont {AC}~\bibnamefont {Bertuola}},\
  }\bibfield  {title} {\enquote {\bibinfo {title} {Aristarco revisitado},}\
  }\href@noop {} {\bibfield  {journal} {\bibinfo  {journal} {Revista Brasileira
  de Ensino de F{\'\i}sica}\ }\textbf {\bibinfo {volume} {38}} (\bibinfo {year}
  {2016})}\BibitemShut {NoStop}%
\bibitem [{\citenamefont {Freitas}\ \emph {et~al.}(2021)\citenamefont
  {Freitas}, \citenamefont {Santucci},\ and\ \citenamefont
  {Marques}}]{Freitas21}%
  \BibitemOpen
  \bibfield  {author} {\bibinfo {author} {\bibfnamefont {Lucas~V}\ \bibnamefont
  {Freitas}}, \bibinfo {author} {\bibfnamefont {Rafael~M}\ \bibnamefont
  {Santucci}}, \ and\ \bibinfo {author} {\bibfnamefont {Ivo~A}\ \bibnamefont
  {Marques}},\ }\bibfield  {title} {\enquote {\bibinfo {title} {{Reinventando o
  m{\'e}todo de Aristarco}},}\ }\href@noop {} {\bibfield  {journal} {\bibinfo
  {journal} {Revista Brasileira de Ensino de F{\'\i}sica}\ }\textbf {\bibinfo
  {volume} {43}} (\bibinfo {year} {2021})}\BibitemShut {NoStop}%
\bibitem [{Pir()}]{Pires11}%
  \BibitemOpen
  \href@noop {} {\emph {\bibinfo {title} {{}Evolu{\c{c}}{\~a}o das Id{\'e}ais
  da F{\'\i}sica}}}\BibitemShut {NoStop}%
\bibitem [{\citenamefont {Newton}(1980)}]{newton80}%
  \BibitemOpen
  \bibfield  {author} {\bibinfo {author} {\bibfnamefont {Robert~R}\
  \bibnamefont {Newton}},\ }\bibfield  {title} {\enquote {\bibinfo {title}
  {{The sources of Eratosthenes measurement of the Earth}},}\ }\href@noop {}
  {\bibfield  {journal} {\bibinfo  {journal} {Quarterly Journal of the Royal
  Astronomical Society}\ }\textbf {\bibinfo {volume} {21}},\ \bibinfo {pages}
  {379} (\bibinfo {year} {1980})}\BibitemShut {NoStop}%
\bibitem [{\citenamefont {Engels}(1985)}]{engels85}%
  \BibitemOpen
  \bibfield  {author} {\bibinfo {author} {\bibfnamefont {Donald}\ \bibnamefont
  {Engels}},\ }\bibfield  {title} {\enquote {\bibinfo {title} {{The length of
  Eratosthenes' stade}},}\ }\href@noop {} {\bibfield  {journal} {\bibinfo
  {journal} {The American Journal of Philology}\ }\textbf {\bibinfo {volume}
  {106}},\ \bibinfo {pages} {298--311} (\bibinfo {year} {1985})}\BibitemShut
  {NoStop}%
\bibitem [{\citenamefont {Pinotsis}(2006)}]{pinotsis06}%
  \BibitemOpen
  \bibfield  {author} {\bibinfo {author} {\bibfnamefont {Antonios~D}\
  \bibnamefont {Pinotsis}},\ }\bibfield  {title} {\enquote {\bibinfo {title}
  {{The significance and errors of Erathosthenes' method for the measurement of
  the size and shape of the Earth's surface}},}\ }\href@noop {} {\bibfield
  {journal} {\bibinfo  {journal} {Journal of Astronomical History and
  Heritage}\ }\textbf {\bibinfo {volume} {9}},\ \bibinfo {pages} {57--63}
  (\bibinfo {year} {2006})}\BibitemShut {NoStop}%
\bibitem [{\citenamefont {{NASA Space Science Data Coordinated
  Archive}}(2021)}]{Nasa}%
  \BibitemOpen
  \bibfield  {author} {\bibinfo {author} {\bibnamefont {{NASA Space Science
  Data Coordinated Archive}}},\ }\href
  {https://nssdc.gsfc.nasa.gov/planetary/factsheet/earthfact.html} {\enquote
  {\bibinfo {title} {{Earth Fact Sheet}},}\ } (\bibinfo {year}
  {2021})\BibitemShut {NoStop}%
\bibitem [{\citenamefont {Wikipédia}(2021)}]{Wikipedia}%
  \BibitemOpen
  \bibfield  {author} {\bibinfo {author} {\bibnamefont {Wikipédia}},\ }\href
  {https://pt.wikipedia.org/w/index.php?title=Raio_terrestre&oldid=61691241}
  {\enquote {\bibinfo {title} {Raio terrestre --- wikipédia{,} a enciclopédia
  livre},}\ } (\bibinfo {year} {2021}),\ \bibinfo {note} {[Online; accessed
  23-julho-2021]}\BibitemShut {NoStop}%
\bibitem [{\citenamefont {Hestenes}(1992)}]{Hestenes92}%
  \BibitemOpen
  \bibfield  {author} {\bibinfo {author} {\bibfnamefont {David}\ \bibnamefont
  {Hestenes}},\ }\bibfield  {title} {\enquote {\bibinfo {title} {{Modeling
  games in the Newtonian world}},}\ }\href@noop {} {\bibfield  {journal}
  {\bibinfo  {journal} {American Journal of Physics}\ }\textbf {\bibinfo
  {volume} {60}},\ \bibinfo {pages} {732--748} (\bibinfo {year}
  {1992})}\BibitemShut {NoStop}%
\bibitem [{\citenamefont {{LADIF-UFRJ}}()}]{ladif}%
  \BibitemOpen
  \bibfield  {author} {\bibinfo {author} {\bibnamefont {{LADIF-UFRJ}}},\ }\href
  {https://ladif.if.ufrj.br} {\enquote {\bibinfo {title} {{ Museu Interativo da
  Física da Universidade Federal do Rio de Janeiro}},}\ }\BibitemShut
  {NoStop}%
\bibitem [{\citenamefont {Dickey}(1995)}]{dickey95}%
  \BibitemOpen
  \bibfield  {author} {\bibinfo {author} {\bibfnamefont {Jean~O.}\ \bibnamefont
  {Dickey}},\ }\bibfield  {title} {\enquote {\bibinfo {title} {Earth rotation
  variations from hours to centuries},}\ }\href {\doibase
  10.1017/S1539299600010339} {\bibfield  {journal} {\bibinfo  {journal}
  {Highlights of Astronomy}\ }\textbf {\bibinfo {volume} {10}},\ \bibinfo
  {pages} {17–44} (\bibinfo {year} {1995})}\BibitemShut {NoStop}%
\bibitem [{\citenamefont {Gross}(2007)}]{gross07}%
  \BibitemOpen
  \bibfield  {author} {\bibinfo {author} {\bibfnamefont {Richard~S}\
  \bibnamefont {Gross}},\ }\bibfield  {title} {\enquote {\bibinfo {title}
  {Earth rotation variations-long period},}\ }\href@noop {} {\bibfield
  {journal} {\bibinfo  {journal} {Treatise on geophysics}\ }\textbf {\bibinfo
  {volume} {3}},\ \bibinfo {pages} {239--294} (\bibinfo {year}
  {2007})}\BibitemShut {NoStop}%
\end{thebibliography}%

\end{document}